# Neutron Diffraction Measurements and First Principles Study of Thermal Motion of Atoms in Select $M_{n+1}AX_n$ and Binary $MX$ Transition Metal Carbide Phases

Nina J. Lane,[1,*] Sven C. Vogel,[2] Gilles Hug,[3] Atsushi Togo,[4] Laurent Chaput,[5] Lars Hultman,[6] Michel W. Barsoum[1]

[1]*Department of Materials Science and Engineering, Drexel University, Philadelphia, PA 19104, USA*
[2]*Los Alamos Neutron Science Center, Los Alamos National Laboratory, Los Alamos, NM 87545, USA*
[3]*Laboratoire d'Etude des Microstructures, ONERA-CNRS, 92322 Châtillon Cedex, France*
[4]*Department of Materials Science and Engineering, Kyoto University, Sakyo, Kyoto 606-8501, Japan*
[5]*Institut Jean Lamour, UMR CNRS 7198, Université de Nancy, France*
[6]*Department of Physics, Chemistry and Biology, Linkoping University, SE-581 83 Linkoping, Sweden*

[*]Corresponding author: lane@drexel.edu

## Abstract

Herein, we compare the thermal vibrations of atoms in select ternary carbides with the formula $M_{n+1}AX_n$ ("MAX phases", $M$ = Ti, Cr; $A$ = Al, Si, Ge; $X$ = C, N) as determined from first principles phonon calculations to those obtained from high-temperature neutron powder diffraction studies. The transition metal carbides TiC, TaC, and WC are also studied to test our methodology on simpler carbides. Good qualitative and quantitative agreement is found between predicted and experimental values for the binary carbides. For all the MAX phases studied – $Ti_3SiC_2$, $Ti_3GeC_2$, $Ti_2AlN$, $Cr_2GeC$ and $Ti_4AlN_3$ – density functional theory calculations predict that the $A$ element vibrates with the highest amplitude and does so anisotropically with a higher amplitude within the basal plane, which is in line with earlier results from high-temperature neutron diffraction studies. In some cases, there are quantitative differences in the absolute values between the theoretical and experimental atomic displacement parameters, such as reversal of anisotropy or a systematic offset of temperature-dependent atomic displacement parameters. The mode-dependent Grüneisen parameters are also computed to explore the anharmonicity in the system.

## I. Introduction

The layered ternary $M_{n+1}AX_n$ ceramics (where $M$ is an early transition metal, $A$ is an A-group element mostly from IIIA to VIA, $X$ is C or N, and $n$ = 1, 2, or 3, resulting in 211, 312 and 413 stoichiometries) have received considerable attention recently.[1-2] These materials, also known as MAX phases, crystallize in the space group $P6_3/mmc$ and contain nanolaminated layers that lead to an unusual combination of properties, including high thermal and electrical conductivities, relatively low Vickers hardness values, exceptional damage tolerance and thermal shock resistance, thermal stability, high stiffness and excellent damping capabilities.[1-2]

The focus of this study is the role of atomic thermal vibrations, or the displacement of atoms from their equilibrium positions, as a function of temperature. Atomic motion is central to many properties and is especially important in considering high temperature damping and transport properties, both electrical and thermal.  This work is a continuation of our work aiming to understand the thermal properties of the MAX phases, primarily using high temperature neutron diffraction (HTND). Using this technique, we reported on the thermal properties of $Ti_3SiC_2$,[3-4] $Ti_3GeC_2$,[3] $Ti_2AlN$,[5] $Cr_2GeC$,[5] and $Ti_4AlN_3$.[6]  These results showed that the A-group elements (Si, Ge, and Al in these cases) vibrate with the highest amplitude, acting as "rattlers" due to their weak bonds *relative* to the stronger *M-X* bonds. It is this rattling effect that is believed to be responsible for the low phonon conductivities of the MAX phases comprised of elements heavier than Al, despite their high specific stiffness values and high Debye temperatures. [1,7-8]

In a first principles study of the thermal properties of the 312 MAX phases $Ti_3SiC_2$, $Ti_3AlC_2$, and $Ti_3GeC_2$ by Togo *et al.* in 2010, it was found that the unusual low-frequency phonon states are likely due to the high-amplitude atomic vibrations of the $A$ elements, Si, Al and Ge respectively.[9]    It was also found that the corresponding atomic motions for these low-frequency bands at the K point in the Brillouin zone can be represented as rotations of the A-group atoms around their average positions. At the M point, the phonon modes can be represented as transverse oscillations of the *A* and Ti atoms parallel to one another within the basal planes.  Furthermore, the atomic motion of the *A* and Ti atoms at these phonon modes were shown to be synchronized. The shapes of the localized bands were slightly different among the three compounds studied by Togo *et al*., suggesting slightly different correlated motion behavior.

Experimental evidence for this phenomenon came about later that year in a recent HTND study of $Ti_3SiC_2$ and $Ti_3GeC_2$,[3] in which anomalous mean bond lengths determined from Rietveld analysis of the time-of-flight data were observed in $Ti_3GeC_2$ during heating. This anomalous bond length behavior was reminiscent of that observed in quartz by Tucker *et al.* in 2001,[10-11] where the apparent decrease in bond lengths with increasing temperatures was attributed to a difference between the instantaneous and average atomic positions due to rigid unit modes (RUMs).[12] While it is not believed that these RUMs exist in the MAX phases – since typically they occur when tetrahedra are corner-sharing while in the MAX phases the octahedra are edge-sharing – it was postulated that other high-temperature phenomena may have caused a difference between the instantaneous and average bond lengths.   Using the experimental anisotropic atomic displacement parameters (ADPs) to estimate the atom positions during thermal motion - and from those the instantaneous interatomic distances between atoms during their presumed synchronized motion - it was shown that the correlated atomic motion modeled by Togo *et al,* [9] discussed above, could explain the results. While $Ti_3SiC_2$ did not show anomalous bond expansions, it was also shown that the correlated motion indicated by the anisotropic ADPs, which were different than those in $Ti_3GeC_2$ in amplitude and direction, would not lead to differences between instantaneous and space-averaged interatomic distances.

Another clue may come from results from resonant ultrasound spectroscopy (RUS) experiments, which showed that ultrasonic attenuation (damping) increases dramatically at a characteristic temperature in some MAX phases when they are heated.[13] For $Ti_3SiC_2$ and most of the other MAX phases studied, the temperature at which this occurred was close to their brittle-to-plastic transition temperature; for $Ti_3GeC_2$, which was the only exception, this temperature was significantly lower. No explanation for this effect is currently available, but defects such as Ge vacancies or a 2-D "melting" of the Ge layers were suggested and may be related to the aforementioned differences indicated by the HTND study.[3]

By now it is clear that there is a phenomenon – most likely related to correlated thermal motion – occurring in bulk $Ti_3GeC_2$ at high temperatures. The increased ultrasonic attenuation at unusually low temperatures from the RUS experiments,[13] as well as the abnormal bond lengths observed by Rietveld analysis of HTND data[3] have provided the first indications, along with the theoretical studies through first principles phonon calculations.[9] Any combination of effects including vacancies, microstructure, secondary phases, preferred orientation, and/or other defects may also play a role. These comments notwithstanding, the reasons for the differences between $Ti_3GeC_2$ and $Ti_3SiC_2$ are still unclear. Further experimental and theoretical work is needed to fully understand the nature and consequences of the proposed correlated atomic motion in the MAX phases. As far as theoretical work, this phenomenon is heavily dependent on the phonon spectrum and therefore first principles phonon calculations serve as a useful tool in investigating the nature of these vibrations. While the phonon spectrum has been used to investigate the thermal expansion and heat capacities of $Ti_3GeC_2$, $Ti_3AlC_2$ and $Ti_3GeC_2$ in Ref. [9], no relationship to the actual atomic displacements was shown. Previously, first principles phonon calculations have been used to predict the thermal atomic displacements in other systems including $SiO_2$,[14-15] $MgB_2$,[16] skutterdites,[17-18] CoO,[19] $NaAlH_4$,[20] and gold nanoparticles.[21] To our knowledge, the mean-squared atomic displacements have not been calculated from first principles phonon calculations for any carbides or nitrides.

Herein, we show that the mean-squared ADPs can be directly calculated and compared to the experimentally determined ones. We report on the temperature dependence of the ADPs of a number of MAX phases based on the phonon spectrum determined from first principles calculations based on density functional theory (DFT).

Titanium carbide, TiC, and tungsten carbide, WC, are also studied with first principles calculations and HTND for benchmarking and as validation for our experimental and theoretical methodologies. TiC is chosen for its similarity in chemistry to the MAX phases studied herein and WC is studied to test our methodology on a hexagonal system. WC crystallizes in a hexagonal structure with space group $P\bar{6}m2$. Both TiC and WC were measured on the same neutron diffractometer as the MAX phases, with the same data refinement strategy. Tantalum carbide, TaC, is also studied with first principles phonon calculations for comparison with another recent HTND paper in which its ADPs were reported in order to evaluate our results against neutron diffraction data from another diffractometer.[22]

## II.    Methods

### i.    *Computational details*

For phonon calculations, 2x2x1 supercells were used, which consisted of 24, 32, and 48 atoms for the 211, 312, and 413 phases, respectively. For TaC and TiC, 2x2x2 supercells consisting of 64 atoms were used; for WC, a 2x2x2 supercell with 8 atoms was used. The DFT calculations were performed using the projector-augmented wave (PAW) [23] method, as implemented in the VASP code. [21-22] The exchange-correlation function used was the Perdew-Burke-Ernzerhof (PBE) generalized gradient approximation (GGA). [24] The plane-wave cutoff was set to 500 eV, and the total energy was converged to $10^{-8}$ eV with a $\Gamma$-centered k-point grid of 6x6x4.

Real-space force constants in the supercells were calculated using density functional perturbation theory (DFPT) [25] implemented in the VASP code. The frequencies were calculated from the force constants using the phonopy code. [26-27]

The displacements of atoms from their equilibrium positions, $\vec{u}_\tau(t)$, are written in terms of annihilation and creation operators as:

$$\vec{u}^\alpha(\tau l, t) = \frac{1}{\sqrt{m_j N}} \sum_{q,p} \exp(i\vec{q} \cdot r(\tau l)) \vec{e}_p^\alpha(\tau, q) \sqrt{\frac{\hbar}{2\omega_p(q)}} \left( a_p(q) \exp(-i\omega_p(q)t) + a_p^\dagger(-q) \exp(i\omega_p(q)t) \right)$$

(1)

where $\tau$ and $l$ indices refer to the summation over the atoms and unit cells in the periodic crystal, respectively, and $r(\tau l)$ is the atomic position of atom $\tau$ in the $l^{th}$ unit cell. $\alpha$ is the Cartesian component, $t$ is time, $q$ is the wave vector, $p$ is the band index, and $\vec{e}_p^\alpha(q)$ are the eigenvectors of the dynamical matrix. Using the completeness and commutations relations for creation and annihilation operators, for any given atom $\tau$ we obtain the average displacement as an expectation value,

$$\left\langle \left| \vec{u}_\tau^\alpha \right| \right\rangle^2 = \left\langle \vec{u}_\tau^\alpha \cdot \vec{u}_\tau^\alpha \right\rangle = \frac{1}{M_\tau N} \sum_{q,p} \vec{e}_p^\alpha(q) \cdot \vec{e}_p^\alpha(-q) \frac{\hbar}{2\omega_p(q)} \left( 1 + 2n_p(q) \right).$$

(2)

Here $\omega_p(q)$ is the phonon frequency, and $n_p(q)$ is the phonon population of mode $(q,p)$:

$$n_p(q) = \frac{1}{\exp\left(\dfrac{\hbar \omega_p(q)}{k_B T}\right) - 1} \quad .$$

(3)

$\left\langle |\vec{u}^{\alpha}| \right\rangle^2$ estimates the mean-squared displacement from its equilibrium position in a given direction. Experimentally the mean-squared displacements are represented as $U_{ij} = \left\langle u_{\tau}^i \right\rangle \left\langle u_{\tau}^j \right\rangle$ ($i,j$=1,2,3) and $U_{11}$, $U_{22}$ and $U_{33}$ are the ADPs in the $x$, $y$ and $z$ directions, respectively. Note that given hexagonal symmetry, $U_{11} = U_{22}$.

To explore the effect of vacancies on thermal motion, the ADPs are also calculated for a 2x2x1 $Ti_3GeC_2$ supercell with one vacant Ge site, representing a material with 12.5% ordered vacancies. The break in symmetry results in 35 single displacements and the frozen phonon method was used to compute the forces induced by finite displacement through the Hellmann-Feynman theorem. The frequencies are calculated from the force constants using the phonopy code, and the temperature-dependent $u_{ij}$ values are calculated from Eqs. 1-3.

The mode-dependent Grüneisen parameters are also calculated from first principles phonon calculations at three different volumes to explore anharmonic contributions. For each phonon (wave vector $q$ and band $p$), the Grüneisen parameter, $\gamma$ - which expresses the volume dependence of mode frequency - is calculated by computing the phonon frequencies at three different volumes and then using the following approximation:

$$\gamma_p(q) = -\frac{V}{\omega_p(q)} \frac{d\omega_p(q)}{dV} \simeq -\frac{V}{2(\omega_p(q))^2} \left\langle e_p(q) \left| \frac{\Delta D(q)}{\Delta V} \right| e_p(q) \right\rangle$$

(4)

where V is the cell volume and $\omega_p(q)$ is the phonon frequency of the mode.

## ii.    *Experimental Details*

The HTND experiments for TiC, WC, $Ti_2AlN$, $Cr_2GeC$, $Ti_3GeC_2$ and $Ti_3SiC_2$ were conducted on the High-Pressure Preferred Orientation neutron diffractometer (HIPPO)[23-24] at the Lujan Neutron Scattering Center, Los Alamos National Laboratory. Information about the experiments and samples can be found in Ref. [3] for $Ti_3SiC_2$ and $Ti_3GeC_2$ and in Ref. 6 for $Ti_2AlN$ and $Cr_2GeC$.

In contrast to the bulk MAX phase samples used, commercially obtained powders were used for TiC (Sigma Aldrich, $\leq 4$ μm powder, $\geq 95\%$ purity) and WC (Alfa Aesar, 99% purity, -100+270 mesh powder). Powder samples were placed in a 9 mm diameter, 0.15 mm wall thickness vanadium, V, holder, mounted in an ILL-type high-temperature vacuum furnace with a V-setup, and heated at a rate of 20 $^o$C/minute. Time-of-flight data were collected at room temperature and then every 100 $^o$C starting at 100 $^o$C upon heating up to 1000 $^o$C, and every 200 $^o$C upon cooling. At each temperature, neutrons were detected with 42 detector panels of $^3$He detector tubes arranged on three rings with nominal diffraction angles of 60, 90 120, and 144 degrees. For TiC, the sample was measured at rotation angles of 0, 45, and 90 degrees around the vertical axis to allow for a full texture analysis at each temperature.

The neutron data were analyzed with the Rietveld method using the General Structure Analysis System (GSAS).[26] The script-controlled refinement strategy, implemented by the *gsaslanguage* refinement script language,[27] insured that identical refinement strategies were used on all samples measured on the HIPPO diffractometer. The instrument alignment (DIFC parameter in GSAS) was calibrated for the highest-resolution detector bank (backscattering at 144 degrees) for the lowest temperature runs and fixed for the subsequent runs. Refined parameters were 16 background parameters of GSAS background function #1, lattice parameters of all phases, instrument calibration (only for the first run), peak width, absorption, and thermal motion parameters. For the MAX phases studied in Refs. 6 and [3], phase fractions and lattice parameters of secondary phases and symmetry-constrained atomic positions were also refined.

For $Ti_4AlN_3$, HTND experiments were conducted on the High Flux Isotope Reactor (HFIR) at the HB-4, high-resolution neutron powder diffractometer. Experimental details and sample synthesis information can be found elsewhere.[6] Note that actual chemistry of the $Ti_4AlN_3$ sample was $Ti_4AlN_{2.9}$.

## III. Results and Discussion

We begin our study on the relatively simple systems, TiC and TaC, which have cubic NaCl-type structures (space group *Fm-3m*). For TiC, we compare the temperature dependence of the mean-squared displacements calculated from first principles phonon calculations with those obtained from HTND. For cubic structures, the thermal motion is represented as an isotropic ADP, $U_{iso}$, which is the mean-square of the displacement of an atom in all directions. **Figure 1a** shows that the calculated temperature dependencies of $U_{iso}$ (lines) are close to the experimental values determined from the HTND carried out herein (symbols), lending credibility to our methodology for this cubic binary system. Both prediction and experiment show nearly the same ADPs for both atoms.

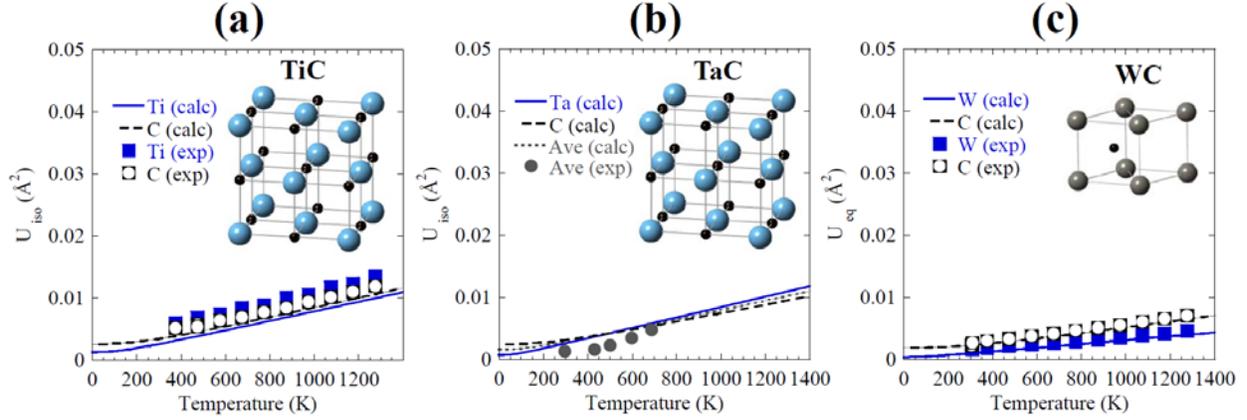

**Fig. 1:** Temperature dependence of mean-squared displacement parameters in select binary carbides showing values calculated from first principles calculations (lines) and experimental values from HTND (markers): (a) $U_{iso}$ for Ti and C atoms in TiC, (b) $U_{iso}$ for Ta and C in TaC, where the markers show the average $U_{iso}$ for Ta and C from Ref. [22] and, (c) $U_{eq}$ for W and C atoms in WC. Insets show the crystal structures for each phase. The full scale in these figures is plotted to coincide with those of all other figures in this paper.

For TaC, we compare the calculated mean-squared displacements with a recent HTND study by Nakamura and Yashima,[22] where the isotropic ADPs were estimated from Rietveld refinement of HTND data on single phase TaC. Since the error bars for $U_{iso}$ were large in that study, the final refinement assumed $U_{Ta} = U_{C}$. Herein the atomic isotropic thermal displacement values are predicted individually for Ta and C, shown in **Fig. 1b** by the blue solid line and black dashed line, respectively. The predicted ADPs for Ta and C are averaged to yield average isotropic ADPs, also shown in **Fig. 1b** (gray dotted line), which are compared with the average $U_{iso}$ values determined from the study in Ref. [22] (gray circles). Note that the $y$-axis limits are chosen to be identical for **Figs. 1-7**, where all ADP plots are shown from 0 to 0.05 Å$^2$ for comparison.

Also noteworthy is the fact that the error bars are largely related to the neutron coherent scattering cross-sections, $\sigma_c$, which are similar for Ta ($\sigma_c = 6$) and C ($\sigma_c = 5.5$);[28] therefore it is appropriate to estimate $U_{iso}$ with an averaged value for the Ta-C system. This comment notwithstanding, the fact that the experimental and calculated ADPs for Ti and C in TiC (**Fig. 1a**) are also similar, despite the fact that their cross-sections are quite different, implies that the symmetry of the crystal is as important.

The predicted and experimental ADP values for TaC are in good agreement, but on this instrument (in the experimental study in Ref. [22]) the temperature dependence shows a more nonlinear behavior at lower temperatures. The extent by which the experimental parameters deviate from the predicted values provide a reference point for the precision of the calculations and the errors involved for simple, single-phase system cubic systems.

To compare the overall amplitudes of vibration of the W and C atoms, the anisotropic ADPs, $U_{ij}$, were converted to equivalent thermal displacement parameters, $U_{eq}$, assuming:

$$U_{eq} = 1/3\left(U_{11} + U_{22} + U_{33} - U_{12}\right).$$  (5)

**Figure 1c** plots the temperature dependence of the $U_{eq}$,'s calculated from first principles phonon calculations for W (solid blue line) and C (dashed black line). The experimental values are represented by blue squares and open circles for W and C, respectively. The agreement between theory and experiment is excellent, with C showing higher amplitudes of vibration in both the predicted and measured results.

To study the directional amplitudes of vibration, we also calculated the anisotropic ADPs, $U_{ij}$, plotted in **Figs. 2a** and **2b** for W and C, respectively. Generally, the predicted values agree well with the experimentally determined ADPs. For C (**Fig. 2b**), the anisotropy is reversed for theory and experiment, *i.e.* $U_{11} > U_{33}$ according to predictions, while the opposite is observed from the HTND experiments. However, in general the calculated values show relatively isotropic behavior ($U_{11}/U_{33} = 1.1$), so this may be an indication of the uncertainty in the degree of anisotropy when the differences between $U_{11}$ and $U_{33}$ are small.

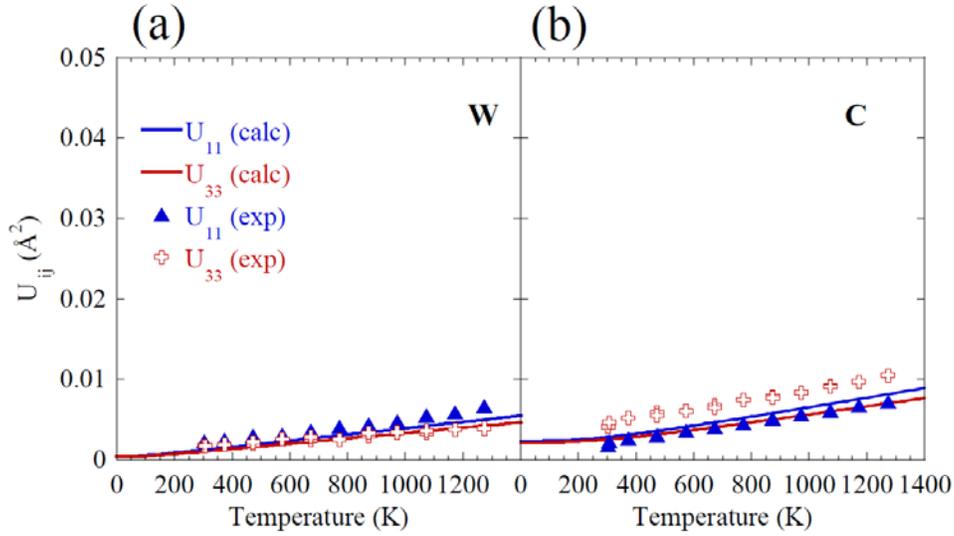

**Fig. 2**: Temperature evolution of anisotropic ADPs $U_{11}$ (blue) and $U_{33}$ (red) of, (a) W and, (b) C in WC. Solid lines show DFT predictions; markers show experimental values determined from HTND.

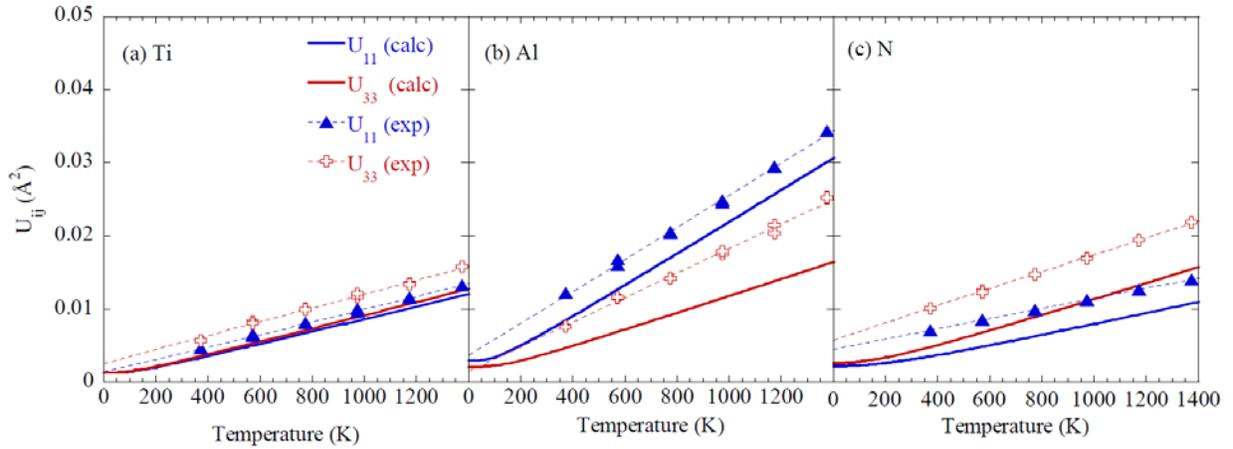

**Fig. 3**: Temperature evolution of anisotropic ADPs. $U_{11}$ (blue) and $U_{33}$ (red) of (a) Ti, (b) Al, and, (c) N atoms in $Ti_2AlN$. Solid lines show DFT predictions; markers show experimental values determined from HTND in Ref. 8.

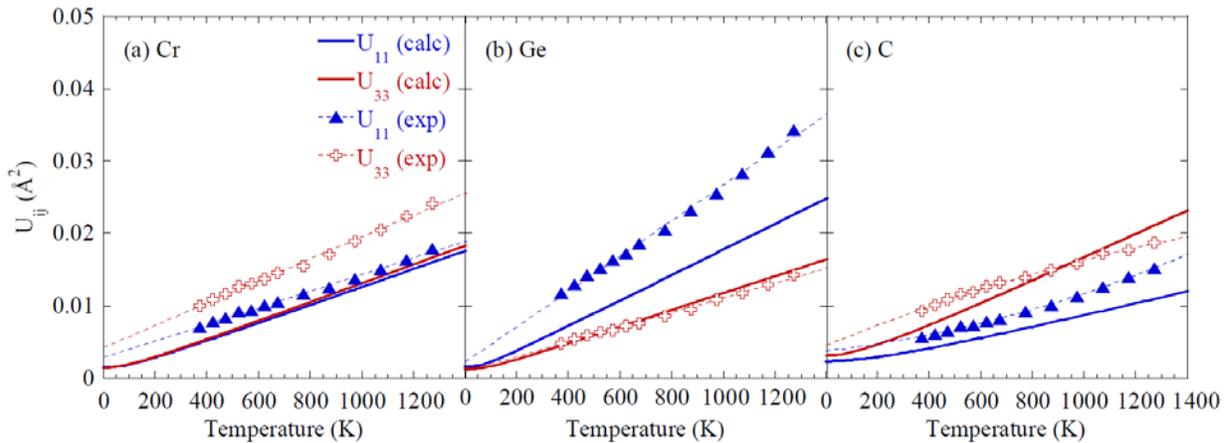

**Fig. 4**: Temperature evolution of anisotropic ADPs $U_{11}$ (blue) and $U_{33}$ (red) of (a) Cr, (b) Ge and, (c) C atoms in $Cr_2GeC$. Solid lines show DFT predictions; markers show experimental values determined from HTND in Ref. 8.

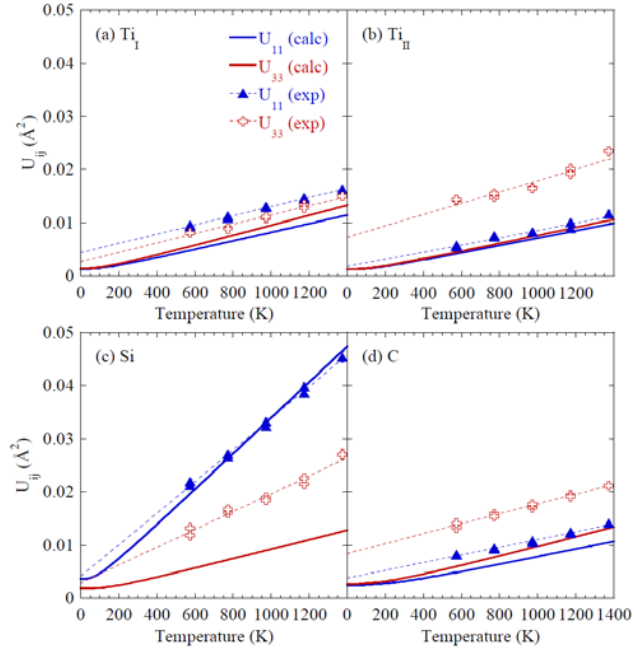

**Fig. 5**: Temperature evolution of anisotropic ADPs $U_{11}$ (blue) and $U_{33}$ (red) of (a) $Ti_I$, (b) $Ti_{II}$, (c) Si and, (d) C atoms in $Ti_3SiC_2$. Solid lines show DFT predictions; markers show experimental values determined from HTND in Ref. 6.

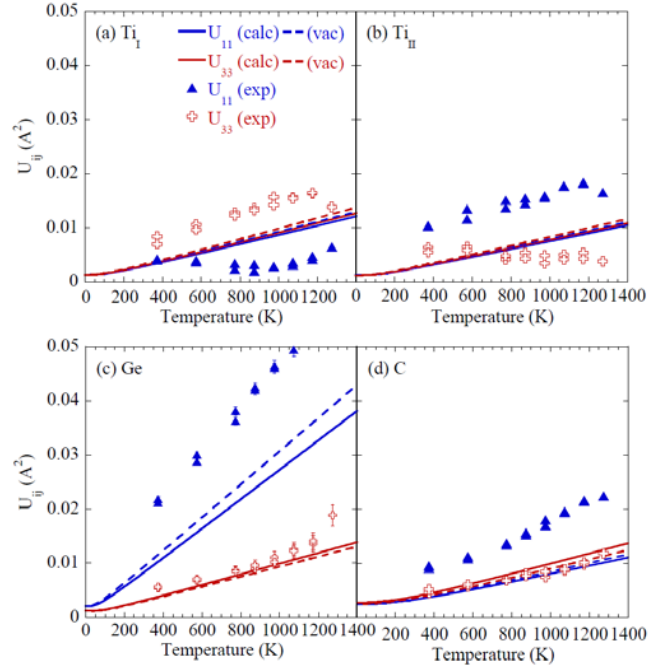

**Fig. 6**: Temperature evolution of anisotropic ADPs $U_{11}$ (blue) and $U_{33}$ (red) of (a) $Ti_I$, (b) $Ti_{II}$, (c) Ge and, (d) C atoms in $Ti_3GeC_2$. Solid lines show first principles predictions for a perfect crystal; dashed lines show DFT predictions for a supercell with 12.5% ordered Ge vacancies; markers show experimental values determined from HTND in Ref. 6.

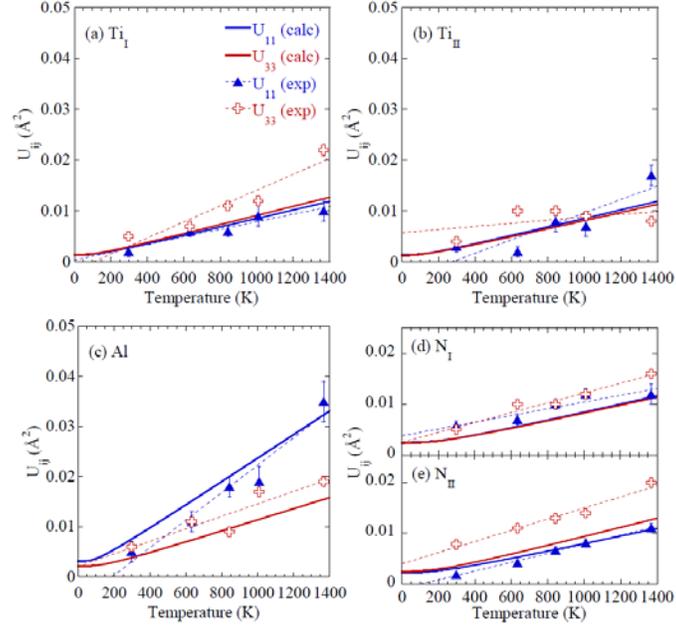

**Fig. 7**: Temperature evolution of anisotropic ADPs $U_{11}$ (blue) and $U_{33}$ (red) of (a) Ti$_{I}$, (b) Ti$_{II}$, (c) Al, (d) N$_{I}$ and, (e) N$_{II}$ atoms in Ti$_4$AlN$_3$. Solid lines show DFT predictions; markers show experimental values determined from HTND in Ref. [6].

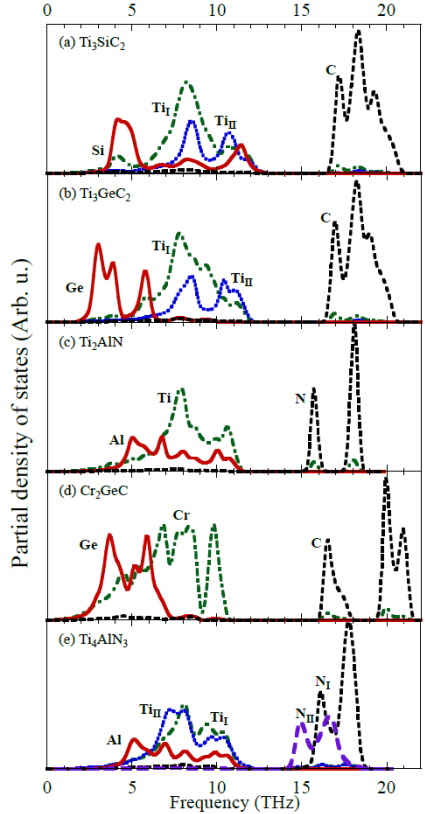

**Fig. 8**: Partial phonon density of states of (a) Ti$_3$SiC$_2$, (b) Ti$_3$GeC$_2$, (c) Ti$_2$AlN, (d) Cr$_2$GeC, and (e) Ti$_4$AlN$_3$.

From the results in **Figs. 1** and **2**, it is evident that, for the binary carbides, experimental and first principles phonon calculations agree reasonably well. The main focus of this work, however, is the MAX phases. F**igures 3**-7 show the ADPs for the five select MAX phases studied herein, including two 211 phases (**Figs. 3-4**), two 312 phases (**Figs. 5-6**), and one 413 phase (**Fig. 7**). Values determined from first principles phonon calculations (solid lines) are shown along with experimental values from Rietveld analysis of neutron time-of-flight data. [6,8-9]

Based on the totality of these results it is reasonable to conclude that for this relatively diverse set of MAX phases, *qualitative* agreement between the calculated and measured ADPs is achieved. In all cases, both the DFT predictions and HTND measurements show that the *A* atom vibrates with the highest amplitude, and the greatest degree of anisotropy, vibrating within the basal plane, *i.e.* $U_{11} > U_{33}$. This is in line with the notion of the A-group elements acting as "rattlers", which is consistent with the low phonon conductivity of many MAX phases.[1,7-8]

With a few exceptions (see below), when the experiments show that $U_{11} > U_{33}$, theory shows the same. This is especially true of the cases where there is a large difference between the $U_{ij}$ values, *i.e.* large anisotropy, mostly of the *A* atoms, such as in Figs. **3b** (Al in Ti$_2$AlC), **4b** (Ge in Cr$_2$GeC), **4c** (C in Cr$_2$GeC), **5c** (Si in Ti$_3$SiC$_2$), **6c** (Ge in Ti$_3$GeC$_2$) and **7c** (Al in Ti$_4$AlN$_3$). Most of the discrepancies, on the other hand, occur for atoms with nearly isotropic thermal motion, *i.e.* for which the differences between the $U_{ij}$ values are small such as in Figs. **4a** (Cr in Cr$_2$GeC), **5a** and **5b** (Ti in Ti$_3$SiC$_2$), **6a and 6b** (Ti in Ti$_3$GeC$_2$), and **7a** (Ti in Ti$_4$AlN$_3$). It should be noted that Ti and Cr are both relatively weak neutron scatterers (for Ti, $\sigma_c$ = 1.485; for Cr, $\sigma_c$= 1.66)[28] and therefore other errors could influence the refined $U_{ij}$ values in the data analysis. Relatively weak scattering power of an element translates to fewer constraints of the structural parameters of those atoms by the experimental diffraction data, manifesting itself for instance as deviations such as larger error bars and more scatter of the values for thermal motion as a function of temperature. For all cases where there is qualitative agreement between predicted and calculated values, the atoms are relatively good neutron scatterers ($\sigma_c > 2$ for Si, Ge, and C).

In general, the anisotropy of the calculated average thermal motion agrees with the HTND data, at least for the *A* atoms. Beyond the agreement in the general trends, there are some slight differences hinting at phenomena that are not accounted for in the harmonic approximation of our Rietveld model. Focusing on the ADPs of A-group elements, since they are the highest, it is interesting that both Al-containing phases (Ti$_2$AlN, **Fig. 3**, and Ti$_4$AlN$_3$, **Fig. 7**) generally show good agreement between theory and experiment, where the anisotropy is well-represented by our calculations with a small offset in magnitude for Ti$_2$AlN. On the other hand, in both Ge-containing phases (Cr$_2$GeC, **Fig. 4,** and Ti$_3$GeC$_2$, **Fig. 6**) $U_{11}$ is experimentally observed to be higher than calculated, while $U_{33}$ shows excellent agreement with first principles calculations. The reverse is true for Si in Ti$_3$SiC$_2$ (**Fig. 5**), where $U_{33}$ determined experimentally is higher than the calculated values, while $U_{11}$ agrees well with first principles calculations.

From the phonon partial density of states (**Fig. 8**), it can be seen that the phonon frequencies of the Ge states are lower than those of the Al and Si states since Ge is heavier. Furthermore, the spread of the Si and Al states (**Figs. 8a**, **c** and **d**) indicates more delocalization since the eigenvectors of the dynamical matrix indicate that the lower-frequency states consist of atomic vibrations within the basal plane, while the higher-frequency states are vibrations perpendicular to the basal plane. In Ti$_3$SiC$_2$, the Si atom vibrating within the basal plane, represented by the localized peak between 3 and 6 THz, is highly localized. Note that in Ref. 9, where a 4x4x1 supercell was used, this band is even narrower. This localized peak manifests itself as a higher degree of anisotropy for Si thermal vibrations than for Ge as determined by first principles calculations (compare **Figs. 5(c)** and **6(c)**). Experimentally this is not observed,[6] which suggests either anharmonic effects, that are not accounted for in our model, discrepancies in our force calculations due to assumptions within DFT, or defects in $Ti_3GeC_2$ (likely Ge vacancies or stacking faults) that may cause the vibrations to shift in amplitude and direction.

Looking more closely at the two 312 phases studied herein ($Ti_3SiC_2$ in **Fig. 5** and $Ti_3GeC_2$ in **Fig. 6**), the phonon calculations predict that Si exhibits the highest amplitude of vibration, while this was not observed from the HTND experiments. This is apparent from the thermal ellipsoid representation of the displacements (**Fig. 9**). While the calculated displacements (right) clearly show that thermal vibrations of the Si atom (top) should be larger than the Ge (bottom), the $U_{ij}$s determined from HTND (left) show that the Ge ellipsoids are more "flattened" and have a higher amplitude within the basal plane.

The reason for this state of affairs is unclear at this time. Sources for the discrepancies observed likely come from experimental conditions that are not taken into account in the first principles phonon calculations herein such as defects (e.g. vacancies and stacking faults), which are most likely in the A layer. Very recent experimental studies on $Ti_3GeC_2$ thin films have suggested samples to be Ge deficient,[29] which was also postulated to be responsible for the high damping measured through RUS.[13] To explore this, the estimated ADPs for $Ti_3GeC_2$ with Ge vacancies are shown in **Fig. 6** as dashed lines. From these results, it is clear that vacancies on the *A* site could lead to a shift in the temperature-dependent ADPs that is more in line with those observed experimentally – most notably, an increase in the $U_{11}$ to $U_{33}$ ratio for Ge. From the HTND experiments, $U_{11}/U_{33}$ for Ge is 3.2, while the ratio predicted by DFT calculations is 2.7 for a perfect crystal and 3.1 for one containing 12.5% vacancies.

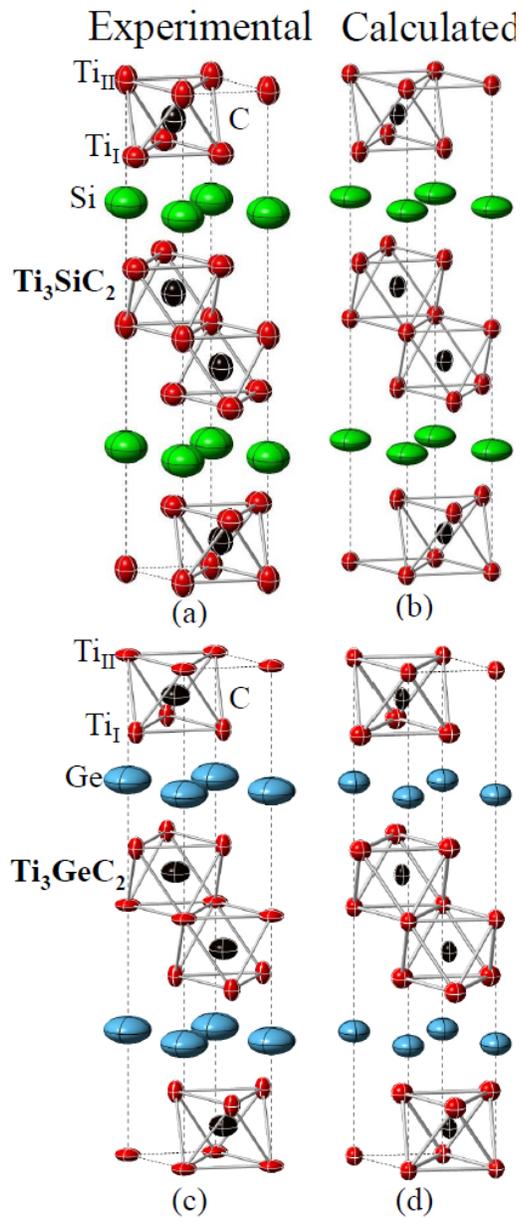

**Fig. 9**: Comparison of 99% probability thermal ellipsoids of atoms in Ti$_3$SiC$_2$ at 1373 K representing (a) experimental and, (b) predicted ADPs; (c) and (d) represent the experimental and predicted thermal atomic displacements, respectively, for Ti$_3$GeC$_2$ at 1273 K.

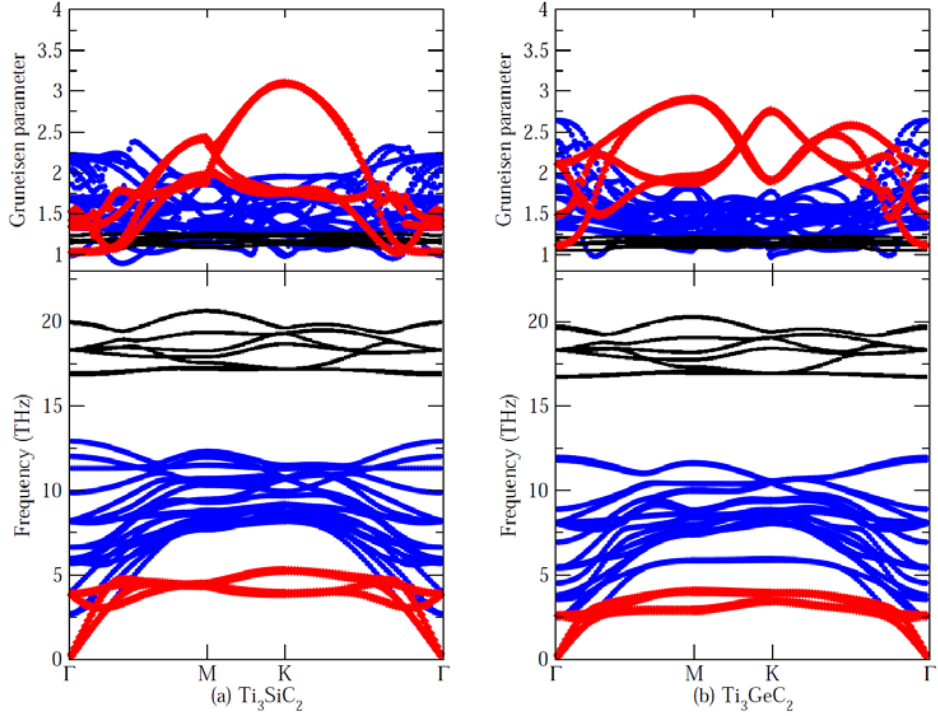

**Fig. 10**: The band structure (bottom) and the mode-dependent Grüneisen parameters (top) for (a) Ti₃SiC₂ and (b) Ti₃GeC₂. The colors in the plots of the Grüneisen parameters correspond to the color scheme of the phonon bands.

The above-mentioned discrepancies may also originate from anharmonicity which is not taken into account in our DFT calculations. To explore this, we calculated the frequency dispersion of the Grüneisen parameter (Eq. 4), which is shown in Figs **10(a)** and **(b)** for Ti₃SiC₂ and Ti₃GeC₂, respectively, along with their phonon band structures. The bands and corresponding Grüneisen parameter curves are color-coded to distinguish between the bands for the Grüneisen parameter dispersions.

For most of the modes in the spectrum, the Grüneisen parameter has a common value, below 2. This gives an average value over the Brillouin zone (1.45 in both cases) that would lead to the conclusion that Ti₃SiC₂ and Ti₃GeC₂ are harmonic compounds to a good approximation. However for the low frequency modes involving the *A* and *M* atoms (in red and green in **Fig. 8**) the Grüneisen parametera are larger. This in turn suggests that the interatomic potential V between A and M atoms is anharmonic because the Grüneisen parameter is proportional to $V'''/V''$. This anharmonicity could contribute to the differences observed. In fact even if usually anharmonicity is evidenced experimentally by a quadratic dependence in the averaged squared displacement, it also modifies the coefficient of the linear term through renormalization of the frequencies. Therefore we have shown that both anharmonicity and the presence of vacancies could play a role on, and have to be considered in, the study of atomic motion.

More work is however needed to understand which one is dominant and to lead to better agreement between theory and experiment. Progress can also be made on the experimental side because there are likely phenomena that cannot be described by the average structure derived from Rietveld analysis of the real-space diffraction patterns, but might be accessible by maximum-entropy methods as described by Izumi *et al.*[30]

## IV. Summary and Conclusions

Herein, we have developed a method for calculating the anisotropic mean-squared atomic thermal displacements through first principles phonon calculations and applied it to select MX and MAX phases. Good qualitative agreement is found between our predictions and HTND experimental results. The frequency dispersions of the Grüneisen parameters for $Ti_3SiC_2$ and $Ti_3GeC_2$ suggest anharmonic interactions between the $M$ and $A$ atoms. The reasons for the quantitative discrepancies between predicted and measured parameters are not totally clear at this time but are most likely related to point defects and/or anharmonic effects.

**Acknowledgements**

This work was partially funded by the Integrated Graduate Education and Research Traineeship (IGERT) under NSF grant number DGE-0654313. This work was also supported by the Foundation for Strategic Research (SSF), Research Council (VR), and Government Strategic Research Area Grant in Materials Science. This work has benefited from the use of the Lujan Neutron Scattering Center at LANSCE, which is funded by the U.S. Department of Energy's Office of Basic Energy Sciences. Los Alamos National Laboratory is operated by Los Alamos National Security LLC under DOE contract DE-AC52-06NA25396. High performance computing resources that have contributed to the research results reported within this paper were provided through the Extreme Science and Engineering Discovery Environment (XSEDE), which is supported by NSF grant number OCI-1053575. The authors gratefully acknowledge the valuable comments of the referee, and helpful discussions with JM Rondinelli.

**Figures:**